\documentclass[preprint]{elsarticle}

\usepackage{amsfonts, amssymb, amsmath, amsthm, mathtools}
\usepackage{physics}
\usepackage{graphicx}
\usepackage{pgfplots}
\pgfplotsset{width=10cm,compat=1.15}
\usetikzlibrary{shapes.arrows}
\usepackage{xcolor}
\usepackage[caption=false]{subfig}

\usepackage{hyperref}
\hypersetup{
    colorlinks=true,
    linkcolor=blue,
    filecolor=magenta,      
    urlcolor=cyan,
    }
\usepackage{cleveref}
\usepackage[inline]{enumitem}
\newcommand{\onlinecite}[1]{\hspace{-1 ex} \nocite{#1}\citenum{#1}}
\usepackage{tikz}
\usetikzlibrary{shapes.geometric, arrows}
\tikzstyle{startstop} = [rectangle, rounded corners, 
minimum width=3cm, 
minimum height=1cm,
text centered, 
draw=black, 
fill=red!30]
\tikzstyle{io} = [trapezium, 
trapezium stretches=true, 
trapezium left angle=70, 
trapezium right angle=110, 
minimum width=3cm, 
minimum height=1cm, text centered, 
draw=black, fill=blue!30]
\tikzstyle{process} = [rectangle, 
minimum width=3cm, 
minimum height=1cm, 
text centered, 
draw=black, 
fill=orange!30]
\tikzstyle{decision} = [diamond,
aspect=2.5,
minimum width=3cm, 
minimum height=1cm, 
text centered, 
draw=black, 
fill=green!30]
\tikzstyle{arrow} = [thick,->,>=stealth]
\usepackage{soul}
\renewcommand\st[1]{}

\journal{Annals of Physics}

\begin{document}

\begin{frontmatter}



\title{Chiral spin liquid state of strongly interacting bosons with a moat dispersion: a Monte Carlo simulation}


\author[inst1]{Chenan Wei}

\author[inst1]{Tigran A. Sedrakyan}

\affiliation[inst1]{organization=
    {Department~of~Physics,~University~of~Massachusetts},
    city={Amherst}, 
    state={Massachusetts},
    postcode={01003},
    country={USA}}

\begin{abstract}
We consider a system of strongly interacting bosons in two dimensions with moat band dispersion which supports an infinitely degenerate energy minimum along a closed contour in the Brillouin zone. The system has been theoretically predicted to stabilize a chiral spin liquid (CSL) ground state. In the thermodynamic limit and vanishing densities, $n\rightarrow 0$, chemical potential, $\mu$, of the uniform CSL state was shown to scale with $n$ as $\mu\sim n^2\log ^2n$. Here we perform a Monte Carlo simulation to find the parametric window for particle density, $n \lesssim \frac{k^2_0}{82 \pi}$, where $k_0$ is the linear size of the moat (the radius for a circular moat), for which the scaling $\sim n^2\log ^2n$ in the equation of state of the homogeneous CSL is preserved. We variationally show that the uniform CSL state is favorable in an interval beyond the obtained scale and present a schematic phase diagram for the system. Our results offer some density estimates for observing the low-density behavior of CSL in time-of-flight experiments with a recently Floquet-engineered moat band system of ultracold atoms in Phys. Rev. Lett. 128, 213401 (2022), and for the recent experiments on emergent excitonic topological order in imbalanced electron-hole bilayers.

\end{abstract}



\begin{keyword} 
Chiral spin liquid \sep Monte Carlo simulation \sep topological order  \sep moat band 
\vspace{1cm}
\begin{center}
{\em Dedicated to the memory of Konstantin B. Efetov    }
\end{center}
\end{keyword}

\end{frontmatter}



\section{Introduction}

Quantum spin liquids (QSLs)\cite{science20,Roderich19,Balents16,PhysRevX.12.021029} are amazing states in quantum many-body condensed matter physics, where the interplay of the lattice gauge theory and topology plays a crucial role\cite{Sachdev18,wang2022emergent,Tigran20,sedrakyan2015spontaneous,Semeghini21}. These states are qualitatively different from the ordinary phases of many electrons and atoms. A characteristic property of QSLs is that they exhibit an absence of breaking of the rotational symmetry and, thus, an absence of long-range ordering. They support the fractionalization of quasiparticle excitations such as anyons, while their analytical description entails the emergence of gauge fields. On the contrary, ordered states lead to massless Goldstone modes, such as spin waves, due to continuous symmetry breaking. Broken continuous symmetries give rise to local order parameters that can describe the zero-temperature ground state of the system.

The concept of emerging gauge fields provided a very natural way to classify and study QSLs in general. Moreover, theoretically, the spin-liquid phase has been linked to the confinement problem in the theory of fundamental interactions. 
The appearance of gauge fields in describing and classifying spin-liquids of lattice systems makes these systems attractive and directly linked to lattice gauge theories. 
The latter theories, originating from particle physics in the context of QCD, provide motivation and a framework for interdisciplinary research toward developing digital and analog quantum simulators and, ultimately, scalable universal quantum computers. In Ref.~\onlinecite{sim20}, two new complementary approaches to studying lattice gauge theories are discussed. First, tensor network methods are presented – a classical simulation approach – applied to studying lattice gauge theories together with some results on Abelian and non-Abelian lattice gauge theories. Then, there have been recent proposals for implementing lattice gauge theory quantum simulators in different quantum hardware, e.g., trapped ions, Rydberg atoms, and superconducting circuits. Finally, the first proof-of-principle trapped ions experimental quantum simulations of the Schwinger model are reviewed.

QSLs, often exhibiting a topological order, feature long-range quantum entanglement that can potentially be exploited to realize robust quantum computation. Ref.~\onlinecite{Semeghini21} used a 219-atom programmable quantum simulator to probe QSL states. In that experiment, arrays of atoms were placed on the links of a kagome lattice, and evolution under the Rydberg blockade created frustrated quantum states with no local order. The onset of a QSL phase of the $Z_2$ toric code type was detected using topological string operators providing direct signatures of topological order and quantum correlations. These observations enable the controlled experimental exploration of topological matter and protected quantum information processing. 
A thorough theoretical analysis of the experiment \cite{Semeghini21} was performed in Ref.~\onlinecite{Lukin22}.  Motivated by these experimental advances, Ref.~\onlinecite{Sachdev22} shows that combining Rydberg interactions and appropriate lattice geometries naturally leads to {\em emergent} $Z_2$ gauge theories endowed with matter fields. 

Ref.~\onlinecite{Tarruell22} reports the quantum simulation of a topological gauge theory by realizing a one-dimensional reduction of the Chern-Simons (CS) theory (the chiral BF theory) in a Bose-Einstein condensate (BEC). This experiment reveals the critical properties of the chiral BF theory: the formation of chiral solitons and the emergence of a self-generated electric field. The results expand the scope of quantum simulation to topological gauge theories and open a route to implementing analogous gauge theories in higher dimensions.

Another efficient framework for the quantum simulation and computation of gauge theories is developed in Ref.~\onlinecite{Wiese}, which is based on quantum link models providing a resource-efficient framework for the quantum simulation and computation of gauge theories.

Ref.~\onlinecite{deconfined22} analyzes a Higgs transition from a U(1) Dirac spin liquid to a gapless $Z_2$ spin liquid. This $Z_2$ spin liquid is relevant to the spin $S = 1/2$ square lattice antiferromagnet. Recent numerical studies have given evidence for such a phase existing in the regime of high frustration between nearest neighbor and next-nearest neighbor antiferromagnetic interactions (the $J_1$-$J_2$ model) appearing in a parameter regime between the vanishing of Néel order and the onset of valence bond solid ordering.

\begin{figure}[!ht]
    \centering
   \begin{minipage}{0.6\linewidth}
    \includegraphics[width=\linewidth]{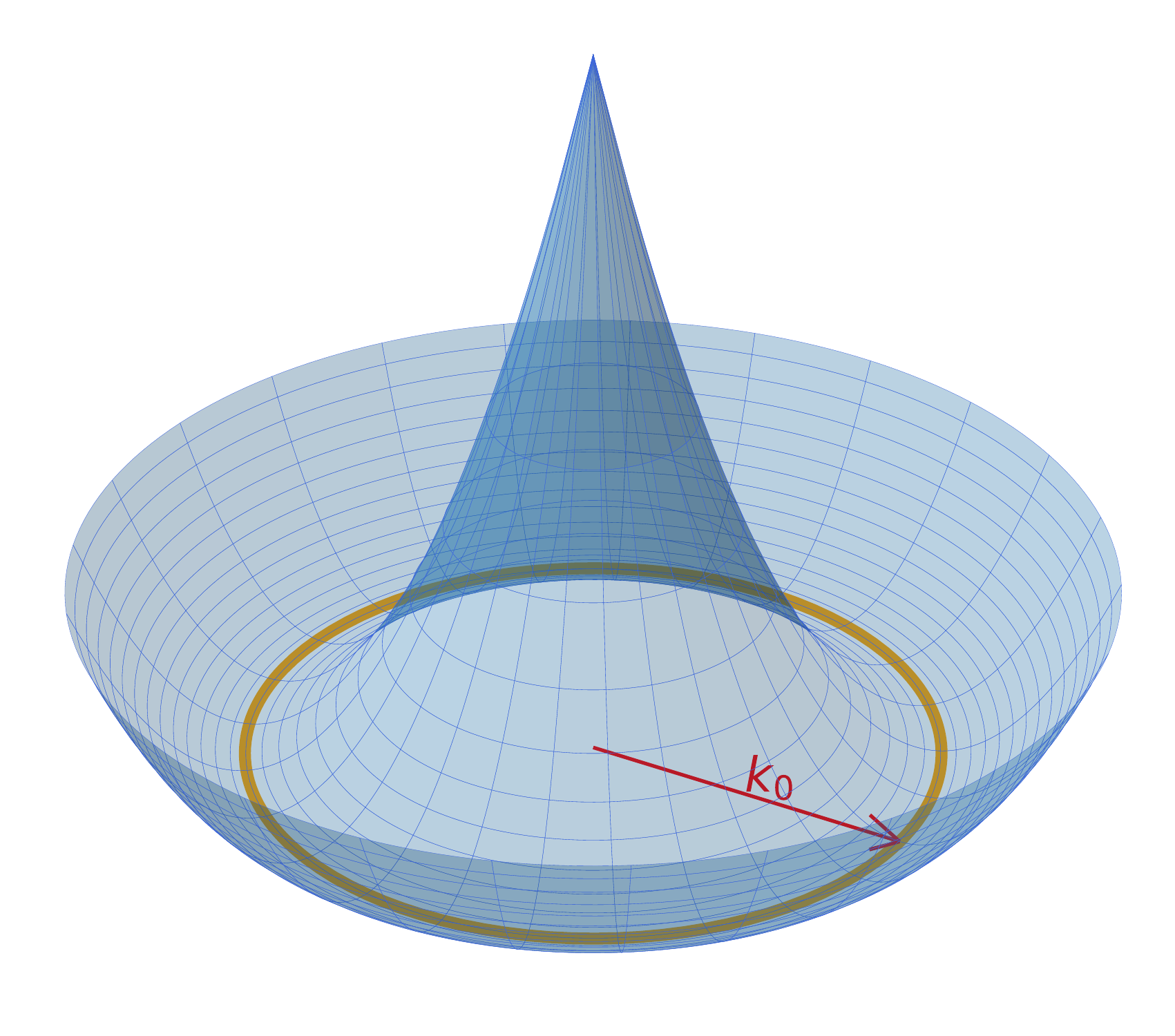}
   \end{minipage}
    \caption{
    Moat dispersion with the chemical potential near the bottom of the band. The radius of the minimal circle is $k_0$.
    }
    \label{fig:moatband}
\end{figure}

Generally, a quantum spin-$1/2$ antiferromagnet can be regarded as a model of hard-core bosons hopping on a lattice. Importantly, once the frustration of quantum spins is strong enough, the lattice dispersion exhibits degeneracies. An intermediate degenerate regime is when the lattice dispersion acquires the moat shape, i.e., the degenerate energy minimum along a closed line in the Brillouin zone, see \cref{fig:moatband}. In this case, as in one dimension (1D), the single-particle density of states diverges at the bottom of the band. In analogy with the 1D Tonks-Girardeau gas\cite{tonks1936the,girardeau1960relationship,lieb1963exact,yang1967some,paredes2004tonks,kinoshita2004observation}, it suggests transforming the quantum  $s=1/2$ spins, or equivalently, hard-core bosons, into spinless fermions, which automatically satisfy the hard-core condition. 
The 1D many-body bosonic wavefunction in the Tonks-Girardeau limit when $g/n\rightarrow \infty$, where $g$ is the interaction parameter and $n$ boson density, is then nullified when bosons reside at the same spatial point. The exact solution of the model in this limit, corresponding to low densities or strong interactions, shows that the wavefunction is in fact equivalent to the absolute value of the fermionic Slater determinant in 1D, which supports bosonic statistics. 

In 2D, such "fermionization" may be achieved with the help of the CS transformation. 
 An example of such statistical transmutation in 2D is the composite boson state in the fractional quantum Hall systems. In the composite boson scenario, Landau quantization keeps the kinetic energy completely degenerate at a given filling factor, and the interactions induce emergent CS fields that attach fluxes to the fermions, leading to statistical transmutation\cite{polyakov1988fermi,jain1989composite,lopez1991fractional,halperin1993theory}.

The chiral spin liquid (CSL) wavefunction within this approach can be obtained from the CS fermion representation of a 2D bosonic system. A widely accepted point of view\cite{kalmeyer1987equivalence,wen1989chiral,3,4} is 
that an interacting spin/boson system can always be formally represented in the form of 
fermions coupled to a CS field.
Within the flux-smearing mean-field theory, fermions in the background magnetic field (typically fully) occupy the lowest Landau level. The wavefunction of the fermion state is given in terms of a Slater determinant. The bosonic wavefunction is obtained upon the multiplication of the Slater determinant by the CS phase, representing the fermionization procedure in the first quantization. Depending on the direction of the field seen by the fermions, the overall CS terms in the ground state wavefunction either cancel out (corresponding to taking the absolute value of the Slater determinant, similarly to 1D) or give rise to a prefactor $\sim$ (CS term)$^2$. The Kalmeyer-Laughlin CSL state is an example of this type. In the second quantization approach, these different situations are obtained by integrating fermions out and looking at the low-energy effective theory. If the CS terms are canceled out, the gauge field dynamics is thus of Maxwell type representing a superfluid. In the second case, it is a variant of the CSL. 

\begin{figure}[!ht]
    \centering
    \begin{minipage}{1.0\linewidth}
    \includegraphics[width=\textwidth]{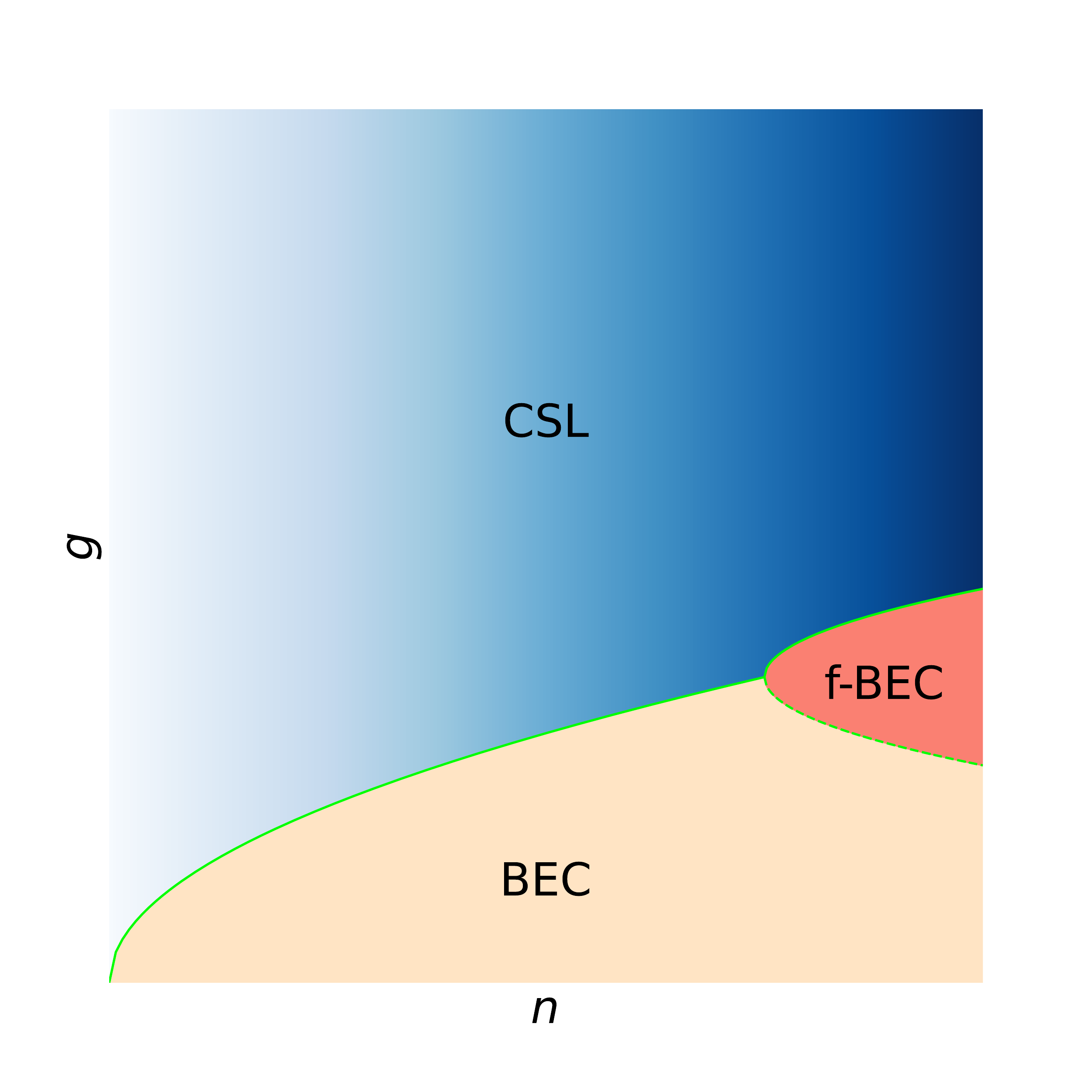}
    \begin{picture}(0,0)
    {\def\unitlength{}
    \put(0.5\textwidth,0.3\textwidth){$U(1)$ symmetry}
    \put(0.55\textwidth,0.27\textwidth){breaking}
    \put(0.6\textwidth,0.32\textwidth){\vector(2,1){0.15\textwidth}}
    \put(0.6\textwidth,0.25\textwidth){\vector(-2,-1){0.05\textwidth}}
    \multiput(0.427\textwidth,0.14\textwidth)(0\textwidth,0.04\textwidth){20}{\line(0,1){0.02\textwidth}}
    \multiput(0.7\textwidth,0.14\textwidth)(0\textwidth,0.04\textwidth){7}{\line(0,1){0.02\textwidth}}
    \put(0.42\textwidth,0.1\textwidth){$n^*_1$}
    \put(0.64\textwidth,0.1\textwidth){$n^*_2 \sim 1.64n^*_1$}
    \put(0.25\textwidth,0.79\textwidth){Non-uniformity increasing}

    \multiput(0.1\textwidth,0.42\textwidth)(0.04\textwidth,0\textwidth){15}{\line(1,0){0.02\textwidth}}
    \put(0\textwidth,0.41\textwidth){$g \sim 1$}

    \begin{tikzpicture}[overlay]
        \node[draw, single arrow,
              minimum height=0.6\textwidth, minimum width=0.1\textwidth,
              single arrow head extend=0.02\textwidth,
              anchor=west, rotate=0] at (0.2\textwidth,0.8\textwidth) {};
    \end{tikzpicture}
    }
    \end{picture}
    \end{minipage}
    \caption{
    Phase diagram of the moat band interacting bosons. The horizontal and vertical axes are particle density $n$ and interacting strength $g$, respectively.
    CSL is the chiral spin liquid state, with a darker color indicating a stronger non-uniformity.
    f-BEC is the fragmented Bose-Einstein condensation that includes a condensate in two diametrically opposite momenta, coherent combinations of macroscopically occupied states with momenta along the moat including the condensates at momenta homogeneously spread along the moat \cite{girvin1987Off,gopalakrishnan2011universal,sur2019metallic,lake2021bose}. BEC is Bose-Einstein condensate at one single momentum on the moat. The density labeled in the figure is $n^*_1=k^2_0 / 82 \pi$, which sets the limit for the equation of state $\mu \sim n^2 \ln^2(n)$ of the uniform CSL.}
    \label{fig:PD}
\end{figure}
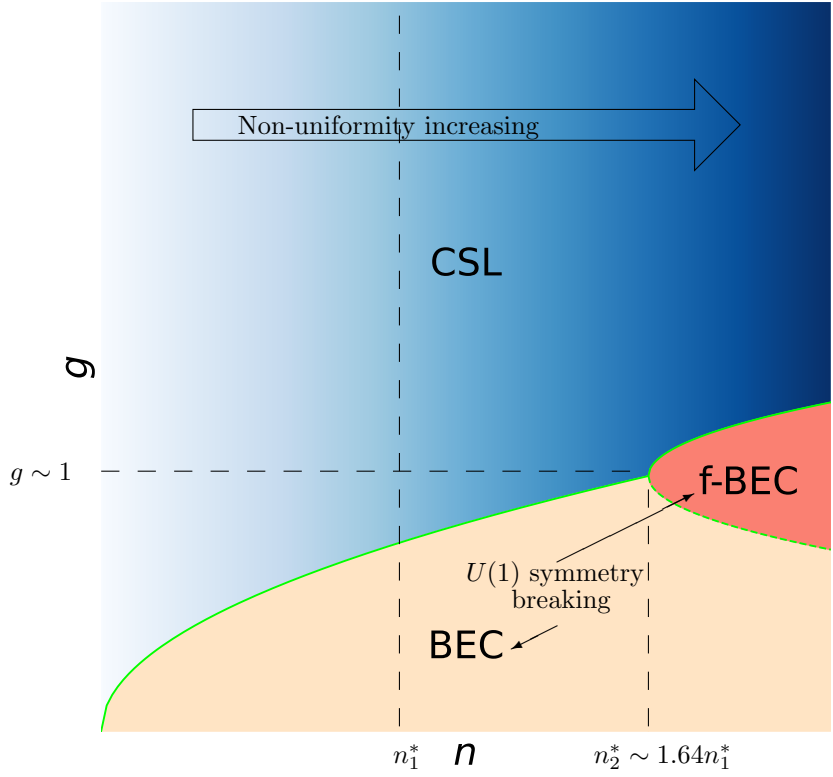

The time-reversal symmetry in a CSL is broken either explicitly at the level of the Hamiltonian\cite{5,hu2016variational,huang2021quantum,zhang2021SU(4),bauer2014chiral}, stabilizing a scalar chirality (like in chiral spin chains supporting the chiral phase with time-reversal symmetry breaking\cite{mkhitaryan2006mean,ambjorn2001integrabel,sedrakyan2001staggered}) or spontaneously\cite{sedrakyan2012compositePRA,sedrakyan2014absence,sedrakyan2015spontaneous,sedrakyan2015statistical,wang2022emergent}. This is an essential aspect of this state of matter, as it can give rise to the appearance of edge states and the ability to store and transmit information in a topologically quantum computation scenario. 
For example, the long-distance physics of the Kalmeyer-Laughlin CSL state can be described by a theory of fermions coupled to the $U(1)$ CS gauge field. After integrating out fermions, the low-energy effective action becomes abelian $U(1)$ CS theory with $K$-matrix being just $K=2$ (or, generally, an even integer $= 2m$). This theory on the torus geometry has a doubly (or $2m$) degenerate ground state. The quasiparticles in the CSL are vortices that support fractionalized charge $e/2$ (or $e/2m$ for $K=2m$) and corresponding anyonic statistics. These excitations are equivalent to Laughlin quasiparticles in the quantum Hall effect, and the vortices of the Kalmeyer-Laughlin state are mutually semions ($\pi$-flux excitations). 

In recent work by one of us (TAS) and collaborators\cite{sedrakyan2012compositePRA,sedrakyan2014absence,sedrakyan2015spontaneous,sedrakyan2015statistical,wang2022emergent}, a  system of strongly interacting bosons in 2D with moat dispersion was considered. It was shown that the degeneracy of the band prevents boson condensation at low densities and/or strong interactions. Furthermore, we showed a formation of a topologically ordered CSL state spontaneously breaking the time-reversal symmetry. This CSL has a bulk gap and chiral gapless bosonic edge excitations\cite{sedrakyan2012compositePRA,sedrakyan2014absence,sedrakyan2015spontaneous,sedrakyan2015statistical}. At low densities, the CSL was shown to be homogeneous, while at higher densities, and particularly at the half-filling of the system the charge density wave (CDW) order or, alternatively, out of plane Ising order for spin-1/2 systems is shown to coexist with the CSL state lowering the overall energy of the CSL state. 
The equation of state corresponding to the homogeneous CSL at low densities, $n\rightarrow0$, implies that the chemical potential scales with the density as 
$\mu_{CSL}\sim n^2 \ln^2n$, see a detailed calculation that considers the fluctuation of the gauge field in\cite{sedrakyan2015statistical}.

The comparison of the chemical potential of the CSL state in interacting bosonic moat band systems with that of the condensate leads to the schematic phase diagram depicted in \cref{fig:PD}. For a regular Bogolyubov-like BEC state with a spontaneously chosen single momentum state on a moat, the equation of state implies the scaling $\mu_{\rm BEC}\sim gn$, where $g$ is the density-density interaction parameter in 2D. The loop-renormalization of the coupling under the assumption of fractionalization of BEC (f-BEC) to two or more diametrically opposite condensates on a moat implies the flow of $g$ to the universal value\cite{gopalakrishnan2011universal} $g\rightarrow g_{\rm univ}\sim n^{1/3}$ at some intermediate densities and interactions leading to the following scaling of the chemical potential $\mu_{\rm f-BEC}\sim n^{4/3}$. Therefore, at sufficiently low particle densities and/or strong interactions\cite{sedrakyan2015spontaneous,sedrakyan2015statistical,wang2022emergent}, the CSL is energetically more favorable.



The realization of the moat band dispersion was also explored in the degenerate quantum gas with Rashba spin-orbit coupling \cite{campbell2011realistic,gopalakrishnan2011universal,ozawa2012stability,galitski2013spin,zhai2015degenerate,berg2012electronic} and bipartite lattices with next-nearest-neighbors hoppings\cite{varney2011kaleidoscope,sedrakyan2014absence,sedrakyan2015spontaneous,wang2017strongly}.
Recently, the moat band was experimentally realized in a system of ultracold atoms in a Floquet-driven optical lattice \cite{bracamontes2022realization}, opening the  possibility of verifying the properties of the variational CSL ansatz experimentally.
In the Floquet setup, by a periodic modulation of optical lattice depth, the initially two separate quadratic bands are adiabatically moved overlappingly and hybridize each other, forming a moat band.
It is, therefore, important to know the numerical range of particle density when the 
$\mu_{\rm CSL}\sim n^2 \ln^2n$
scaling of the low-density
CSL state is stabilized. Since the analytical calculation of $\mu_{\rm CSL}$ is performed only at the low density and thermodynamic limit ($N \to \infty$, $N / V \to 0$)\cite{sedrakyan2015statistical}.

In the present work, the Monte Carlo (MC) calculation is performed to benchmark the low-density result for CSL and find the interval for finite densities for which  $\mu_{\rm CSL}\sim n^2 \ln^2n$. This will help to find the 
CSL state in time of flight experiments for which the momentum distribution will be homogeneous with the width of the peak centered around the moat scaling with $n$ as $\delta k\sim n\log n$. Our Monte-Carlo simulation  indicates $\mu_{\rm CSL} \sim n^2\ln^2n$ relation holds at $n \lesssim \frac{k^2_0}{82 \pi}$. At higher densities, we see that  the equation of state of the CSL state is changing, while the tricritical transition to the BEC and f-BEC states takes place at $g\sim 1$ and densities of the order of $\frac{k^2_0}{50 \pi}$.


\section{Uniform chiral spin liquid state}

There is a direct correspondence between the hard-core lattice boson systems and $S=1/2$ XY lattice magnets. In this way, the BEC state with broken U(1) symmetry translates into a magnetically ordered state of the magnet. 
Here our goal is to study the phase diagram of bosons with strong repulsion, $g>1$, all the way up to the hard-core limit, supporting the moat dispersion (which quite generally describes frustrated magnets on moat lattices) by performing variational Monte Carlo simulations with our accurate wave functions. In particular, we will consider the problem of a moat band populated with bosons with low density.


The system of bosons interacting via Dirac-$\delta$ potential and supporting a moat-like dispersion is described by the Hamiltonian
\begin{equation}
    H = \int d\mathbf{r} \left[ \frac{1}{2M} \Phi^{\dagger}(\mathbf{r}) (|\mathbf{k}| - k_0)^2 \Phi(\mathbf{r}) + \frac{g}{M} [\Phi(\mathbf{r})^{\dagger} \Phi({\mathbf{r}})]^2 \right],
\end{equation}
where $\Phi(r)$ is the bosonic field operator, $\mathbf{k}$ is the momentum operator, $k_0$ is the constant momentum representing the radius of the moat, $M$ is the mass of the boson, and $g$ is dimensionless interacting constant.

The variational ground state many-body wavefunction we consider in the present work is given by\cite{sedrakyan2015statistical}
\begin{equation} \label{PhiB}
    \Phi_B^{(l)} (z_1, z_2, ..., z_N) = \prod_{i<j}^N \frac{z_i - z_j}{|z_i - z_j|} \Psi_F^{(l)} (z_1, z_2, ..., z_N),
\end{equation}
where $z = x + iy$ is the complex variable representing the position of a boson, and $\Psi_F^{(l)}$ is a Slater determinant generated by the single particle Landau wavefunctions corresponding to the fully filled Landau level.
The dependence of wavefunctions on conjugate variables $\bar{z}_1, \bar{z}_2, \cdots, \bar{z}_N$ will be assumed but will not be explicitly written for simplicity throughout the paper. The global phase in front leads to an emergent CS field. Under the mean-field assumption of spatially uniform particle density, $n$, and hence constant strength of the CS field, Landau levels are given by $E = \frac{1}{2M} (\sqrt{(2l+1) M \omega_c} - k_0)^2$, where $\omega_c = B / M = 2\pi n / M$. 
Therefore the energy is minimized at density
\begin{equation} \label{nl}
    n_l = \frac{k_0^2}{2\pi (2l + 1)}, \quad l \in \mathbb{N}.
\end{equation}
The corresponding Slater determinant representing the fully filled Landau level has the form
\begin{equation}\label{PsiF}
    \Psi_F^{(l)} (z_1, z_2, ..., z_N) = \frac{1}{\sqrt{N!}} \mathop{\det_{-l \le m \le N - l - 1}}_{1 \le j \le N} [\chi_m^{(l)} (z_j)].
\end{equation}
The Landau wavefunction of the single particle in the mean-field CS field, $\chi_m^{(l)} (z)$, in the symmetric gauge, reads
\begin{equation}
    \chi_{m}^{(l)}(z) =
    \begin{cases}
    \frac{(-1)^{l} \sqrt{l !}}{l_{B} \sqrt{2 \pi 2^{m}(l+m) !}} \left(\frac{z}{l_{B}}\right)^{m} e^{-\frac{|z|^{2}}{4 l_{B}^{2}}} L_{l}^{(m)} \left[\frac{|z|^{2}}{2 l_{B}^{2}}\right], \quad m \ge 0 \\
    \frac{(-1)^{(l+m)} \sqrt{(l+m) !}}{l_{B} \sqrt{2 \pi 2^{-m} l!}} \left(\frac{\bar{z}}{l_{B}}\right)^{-m} e^{-\frac{|z|^{2}}{4 l_{B}^{2}}} L_{l+m}^{(-m)} \left[\frac{|z|^{2}}{2 l_{B}^{2}}\right], \quad m \le 0 
    \end{cases}.
\end{equation}
Here $ L_{l}^{(m)} (x)$ is the generalized Laguerre polynomial and $l_B = 1/\sqrt{2\pi n_l}$ is the magnetic length. $ L_{l}^{(m)} (x)$ forms a basis of Landau level degenerate subspace at the given density \cref{nl}. The wavefunction \ref{PhiB} preserves the rotational $U(1)$ symmetry and describes the uniform CSL (uniform in the sense that it supports no out-of-plane order).

It is instructive to look at the $l=0$ state when the trial wavefunction becomes
\begin{equation} \label{l0wavefunction}
\begin{aligned}
    \Phi_{B}^{(l=0)}(z_1, z_2, ..., z_N) &\propto \left( \prod_{i<j}^N \frac{z_i - z_j}{|z_i - z_j|} \right) \det V(\{z_j\}) \exp(-\sum_j \frac{|z_j|^{2}}{4 l_{B}^{2}}) \\
    &= \left( \prod_{i<j}^N \frac{z_i - z_j}{|z_i - z_j|} \right)^2 \prod_{i<j}^N |z_i - z_j| \exp(-\sum_j \frac{|z_j|^{2}}{4 l_{B}^{2}}).
\end{aligned}    
\end{equation}
Here $V(\{z_j\})$ is the Vandermonde matrix, the determinant of which is $\det V(\{z_j\}) = \prod_{i<j}^N (z_i - z_j)$. This should be compared with one of the possible f-BEC configurations corresponding to the situation when condensation happens on all the wavevectors along the minima of the moat band. Such a state can be described by the wavefunction written as the modulus of the lowest Landau level wavefunction\cite{girvin1987Off}
\begin{equation} \label{fBECwavefunction}
\begin{aligned}
    \Phi_{\rm f-BEC}^{all}(z_1, z_2, ..., z_N) &\propto \prod_{i<j}^N |z_i - z_j| \exp(-\sum_j \frac{|z_j|^{2}}{4 l_{B}^{2}}).
\end{aligned}    
\end{equation}
Comparing \cref{l0wavefunction,fBECwavefunction}, one can establish the following connection between the two:
\begin{equation}
\Phi_{B}^{(l=0)}(z_1, z_2, ..., z_N) = \left( \prod_{i<j}^N \frac{z_i - z_j}{|z_i - z_j|} \right)^2 \Phi_{\rm f-BEC}^{all}(z_1, z_2, ..., z_N).
\end{equation}
From the field theoretical perspective, this state can be interpreted as arising from the CS gauge theory with a single-valued matrix, $K=2$, after the matter field degrees of freedom in the CSL are integrated out.

In general, when $N \gg 1$, 
\begin{equation} \label{lwavefunction}
\begin{aligned}
    \Phi_{B}^{(l)}(z_1, z_2, ..., z_N) &\propto \left( \prod_{i<j}^N \frac{z_i - z_j}{|z_i - z_j|} \right) \det W(\{z_j\}) \exp(-\sum_j \frac{|z_j|^{2}}{4 l_{B}^{2}}) \\
    &= \left( \prod_{i<j}^N \frac{z_i - z_j}{|z_i - z_j|} \right)^2 \det \frac{W(\{z_j\})}{V(\{z_j\})} \prod_{i<j}^N |z_i - z_j| \exp(-\sum_j \frac{|z_j|^{2}}{4 l_{B}^{2}}) ,
\end{aligned}    
\end{equation}
where the components of matrix $W(\{z_j\})$ are $W_{im}(\{z_j\}) = z_i^{m} L_{l}^{(m)}\left[\frac{|z_i|^{2}}{2 l_{B}^{2}}\right]$. We expect the bosonic part of the many-body state {\em without} the CS phase would lead to other types of f-BEC condensates with a finite condensate fraction on all the momenta along the moat, such as the states discussed in Refs.~\onlinecite{sur2019metallic,lake2021bose}.

\section{Expectation values with uniform CSL states}
\subsection{Variational energy}

In the present section, we will rewrite the energy corresponding to the homogeneous CSL state in an analytical form convenient for MC simulation. 
With the variational wavefunction \cref{PhiB}, the total energy functional is
\begin{equation}
    E_l = \frac{k_0^2}{2M} \bra{\Phi_B^{(l)}((z_1, z_2, ..., z_N))} \sum_i (|k_i| - 1)^2 \ket{\Phi_B^{(l)}((z_1, z_2, ..., z_N))},
\end{equation}
where $k_i = k_{x,i} / k_0 - i k_{y,i}/k_0$ are dimensionless momentum operators.
By permutation symmetry of the arguments of the many-body wavefunction, the per particle energy functional can be computed by
\begin{equation} \label{Esum}
\begin{aligned}
    E_l / N =& \frac{k_0^2}{2M} \bra{\Phi_B^{(l)}((z, z_2, ..., z_N))} (|k| - 1)^2 \ket{\Phi_B^{(l)}((z, z_2, ..., z_N))} \\
    =& \frac{k_0^2}{2M} \left[ \bra{\Phi_B^{(l)}(z, z_2, ..., z_N)} k^{\dagger}k \ket{\Phi_B^{(l)}(z, z_2, ..., z_N)} \right. \\
    & \left. -2 \bra{\Phi_B^{(l)}(z, z_2, ..., z_N)} |k| \ket{\Phi_B^{(l)}(z, z_2, ..., z_N)}
    + 1 \right],
\end{aligned}
\end{equation}
where operator $k = k_x / k_0 - i k_y/k_0$ acts only on the first argument.

As the next step, we introduce the notation $P(z_1)$ as
\begin{equation}
    P(z_1) \equiv \prod_{i<j}^N \frac{z_i - z_j}{|z_i - z_j|},
\end{equation}
for simplicity. Then, the terms in \cref{Esum} can be computed separately.
The first term at the right-hand-side (RHS) becomes equal to
\begin{equation} \label{Ek2}
    \bra{\Phi_B^{(l)}} k^{\dagger}k \ket{\Phi_B^{(l)}}
    = \bra{\Psi_F^{(l)}} (P^* k^{\dagger}k P) + (P^* k P) k^{\dagger}
    + (P^* k^{\dagger} P)k + k^{\dagger}k \ket{\Psi_F^{(l)}}.
\end{equation}
To simplify the second term at the RHS, we use the identity
\begin{equation} \label{absk2int}
    |k| f(z, \bar{z})= \int dw d\bar{w} \frac{1}{2 \pi} \frac{1}{|z-w|} k_w^{\dagger}k_w f(w, \bar{w}).
\end{equation}
The detailed proof of this identity is presented in \ref{apx_proof_absk2int}.
After that, the second term becomes
\begin{equation} \label{Ek}
\begin{aligned}
    & \bra{\Phi_B^{(l)}} |k| \ket{\Phi_B^{(l)}} \\
    =& \frac{1}{2\pi} \int dw d\bar{w} \bra{\Psi_F^{(l)} (z, z_2, ..., z_N)} \frac{1}{|z-w|} \\
    &\left[ (P(z) k^{\dagger}_w k_w P^*(w)) + (P(z) k^{\dagger}_w P^*(w)) k_w \right. \\
    &\left. + (P(z) k_w P^*(w)) k^{\dagger}_w + k^{\dagger}_w k_w \right] \ket{\Psi_F^{(l)} (w, z_2, ..., z_N)},
\end{aligned}
\end{equation}

With the help of \cref{Ek2} and \cref{Ek}, one can rewrite \cref{Esum} in the following form
\begin{equation} \label{El}
\begin{aligned}
    E_l / N =& \frac{k_0^2}{2M} \Big[ 1 + \int dz d\bar{z} \prod_{i=2}^{N} dz_i d\bar{z}_i \Psi_F^{(l)*}(z) \Psi_F^{(l)}(z) \mathcal{A} \\
    & -\frac{1}{\pi} \int dw d\bar{w} \int dz d\bar{z} \prod_{i=2}^{N} dz_i d\bar{z}_i {\frac{e^{-|w-z|}}{2\pi |w-z|}} \Psi_F^{(l)*} (z) \Psi_F^{(l)} (z) \mathcal{B} \Big],
\end{aligned}
\end{equation}
where
\begin{equation}\label{energyFunctional}
\begin{aligned}
    \mathcal{A} =& (P^*(z) k^{\dagger}k P(z)) \\
    &+ \frac{\left[ (P^*(z) k P(z)) k^{\dagger} + (P^*(z) k^{\dagger} P(z))k + k^{\dagger}k \right] \Psi_F^{(l)}(z)}{\Psi_F^{(l)}(z)},
\end{aligned}
\end{equation}
and
\begin{equation}
\begin{aligned}
    \mathcal{B} =& {2\pi e^{|z-w|}}
    \frac{\Psi_F^{(l)} (w)}{\Psi_F^{(l)} (z)} \Big\{ (P(z) k^{\dagger}_w k_w P^*(z)) + \\
    & \frac{\left[  (P(z) k^{\dagger}_w P^*(w)) k_w + (P(z) k_w P^*(w)) k^{\dagger}_w + k^{\dagger}_w k_w \right] \Psi_F^{(l)} (w)}{\Psi_F^{(l)} (w)} \Big\}.
\end{aligned}
\end{equation}
In the above equations, $\Psi_F^{(l)} (w)$ is an abbreviation for $\Psi_F^{(l)} (w, z_2, ..., z_N)$. The normalized wavefunction makes it possible for a probability interpretation. $\mathcal{A}$ and $\mathcal{B}$ are arranged in a way convenient for the rank-1 update for the determinants\cite{ceperly1977monte} and are sampled from the distribution $\Psi_F^{(l)*}(z) \Psi_F^{(l)}(z)$. Therefore, \cref{energyFunctional} is ready for an MC simulation.

\subsection{Scaling of the averaged momentum and variance of momentum}

Before describing our MC simulation results, we will first discuss the analytically expected scaling behavior of $\langle |k| \rangle_B$ and $\langle |k|^2 \rangle_B$ as a function of density $n$, where the notation $\langle \cdots \rangle_B  \equiv \bra{\Phi_B} \cdots \ket{\Phi_B}$ corresponds to the expectation value with respect to the bosonic wavefunction \cref{PhiB}. 
Because of the presence of the CS phase in \cref{PhiB}, the average over the bosonic wavefunction can be written as the average over the fermionic wavefunction where the covariant momentum in introduced elongated with an extra CS gauge potential $A(z) = -i \sum_j (z-z_j) / |z-z_j|^2$, namely, 
\begin{equation}
    \langle |k|^2 \rangle_B = \langle |k - A|^2 \rangle_F.
\end{equation}
Here the $\langle \cdots \rangle_F  \equiv \bra{\Psi_F} \cdots \ket{\Psi_F}$ is the expectation value with respect to the fermionic wavefunction \cref{PsiF}.

The CS gauge potential can be separated into two parts, the mean-field part and the fluctuation
\begin{equation}
\label{covariant}
    |k - A|^2 = |k - \bar{A}|^2 + H_0
\end{equation}
where
the mean field gauge potential is
\begin{equation}
    \bar{A} = -i \sum_{m=-l}^{N-l} \int d z^{\prime} \chi_m^*\left(z^{\prime}\right) \frac{z-z^{\prime}}{\left|z-z^{\prime}\right|^2} \chi_m\left(z^{\prime}\right),
\end{equation}
and the fluctuation is
\begin{equation}
    H_0 = \left[ (\bar{A}^{\dagger} - A^{\dagger}) k + h.c. \right] + ( A^{\dagger} A - \bar{A}^{\dagger} \bar{A} ).
\end{equation}
After averaging \cref{covariant}, the mean-field part gives a constant, $k_0$. Similarly to the derivation presented in the supplementary material of Ref.~\onlinecite{sedrakyan2015statistical}, by expanding the generalized Laguerre polynomial, one can show that
\begin{equation} \label{H0_scaling}
    \langle H_0 \rangle_F \sim C n \ln(n) + D n^2 \ln^2(n).
\end{equation}
Here coefficients $C$ and $D$ are some constants, the precise forms of which are unimportant for our further study. Thus, from here one can conclude that the average $\langle |k|^2 \rangle_B$ depends on density $n$ as
\begin{equation}\label{scaling:k2}
    \langle |k|^2 \rangle_B = k_0^2 + \langle H_0 \rangle_F \sim k_0^2 + C n \ln(n) + D n^2 \ln^2(n).
\end{equation}

Similarly, $\langle |k| \rangle_B$ depends on density $n$ polynomially, \textit{i.e.},
\begin{equation}\label{scaling:k}
    \langle |k| \rangle_B = k_0 + \frac{C}{2k_0} n ln(n) + \frac{D'}{k_0^3} n^2 ln^2(n) + O(n^3 ln^3(n)),
\end{equation}
where $C$ is the same constant as in \cref{H0_scaling} and $D'$ is a new constant. A more detailed derivation of \cref{scaling:k} is given in \ref{apx_proof_scaling:k}. It is instructive to compare the linear terms in \cref{scaling:k2,scaling:k}, whose linearity coefficients $K_2 = C$ and $K_1 = \frac{C}{2k_0}$ are proportional to each other, giving the constant ratio:
\begin{equation} \label{linearTerm}
    \frac{K_2}{k_0 K_1} = 2.
\end{equation}
We will confirm this low-density
 relation by MC simulation in \cref{section:MC}.

\section{MC simulation}\label{section:MC}

\begin{figure}[!ht]
    \centering
  \begin{minipage}{\linewidth}
    \includegraphics[width=\textwidth]{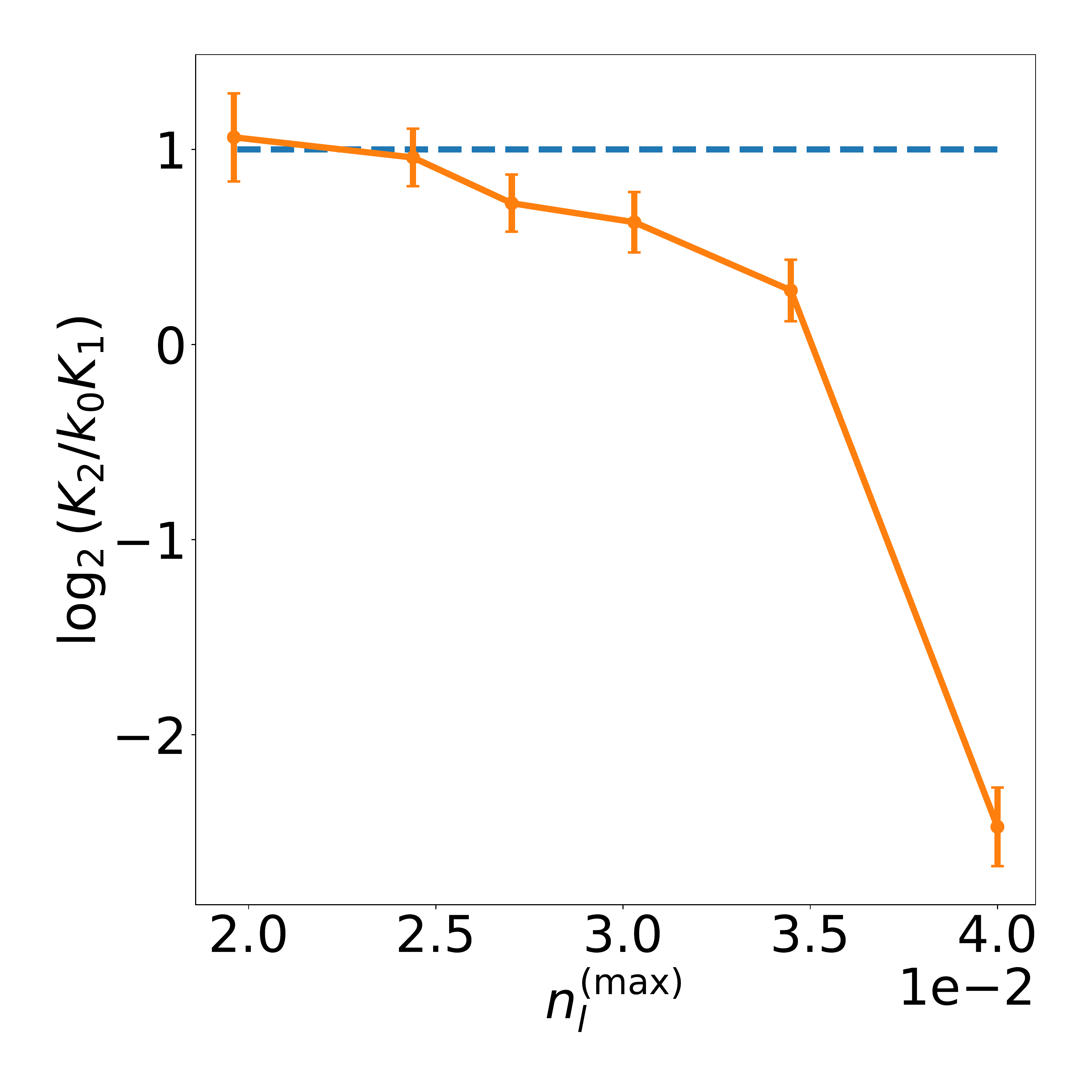}
    \begin{picture}(0,0)
    {\def\unitlength{}
    \put(0.2\textwidth,0.51\textwidth){\includegraphics[height=0.31\textwidth]{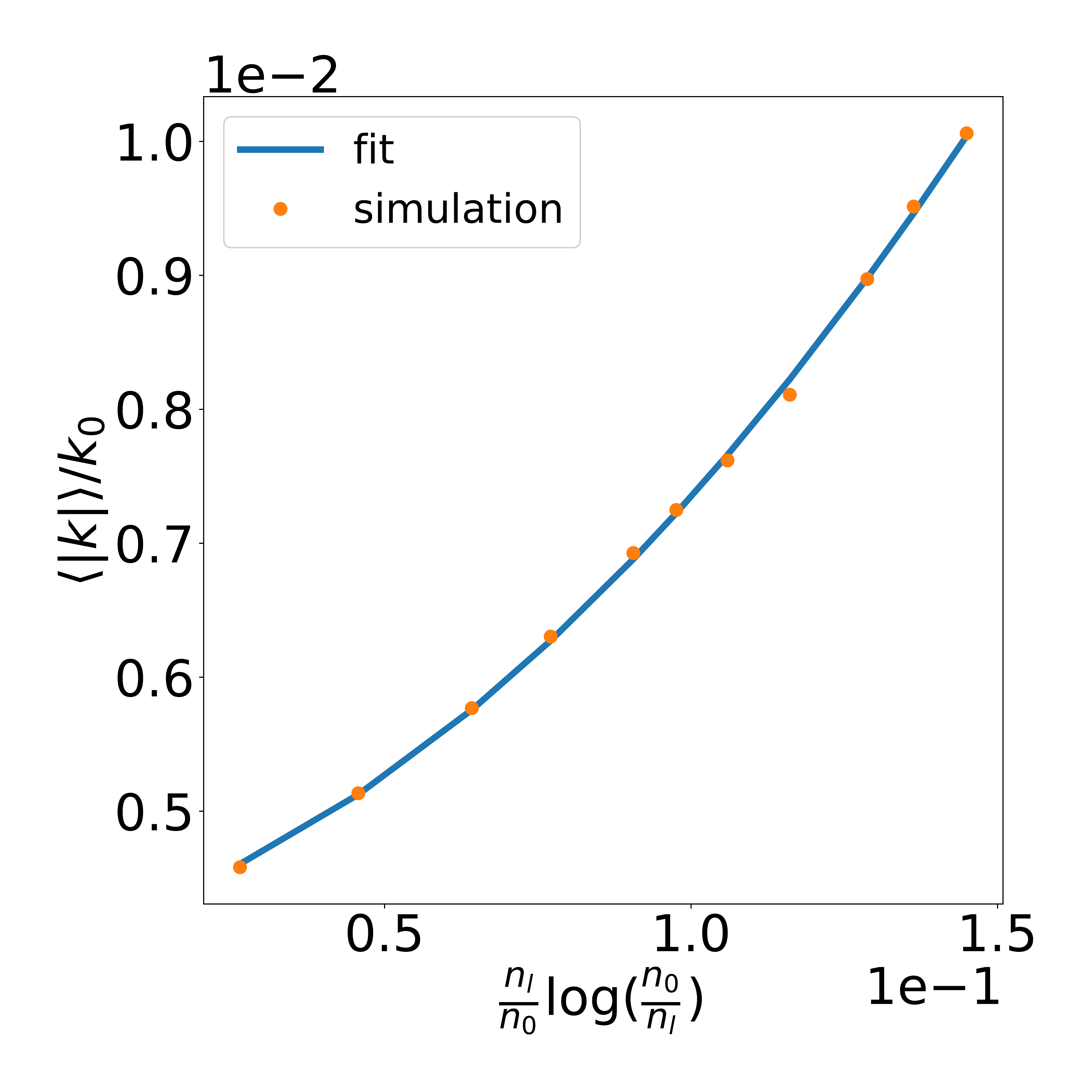}}
    \put(0.2\textwidth,0.21\textwidth){\includegraphics[height=0.31\textwidth]{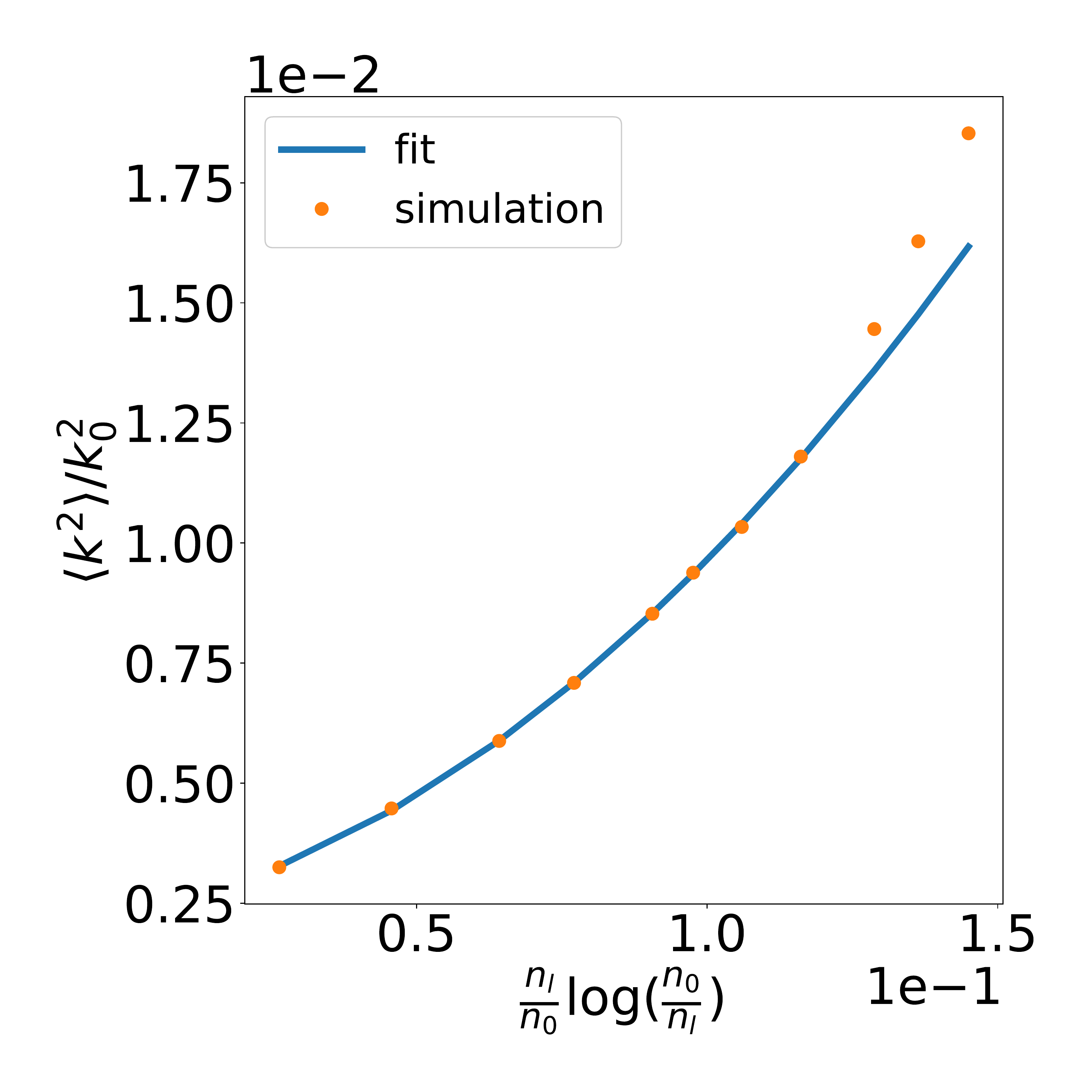}}
    }
    \end{picture}
  \end{minipage}
    \caption{The ratio of the linearity coefficients \cref{linearTerm} at the low density limit. The dashed line works as a guide to the eye. The upper and lower insets separately show the quadratic dependence of averaged $|k|$ and $k^2$.
    }
    \label{fig:ratiok}
\end{figure}

\begin{figure}[!ht]
    \centering
    \begin{minipage}{\linewidth}
    \includegraphics[width=\textwidth]{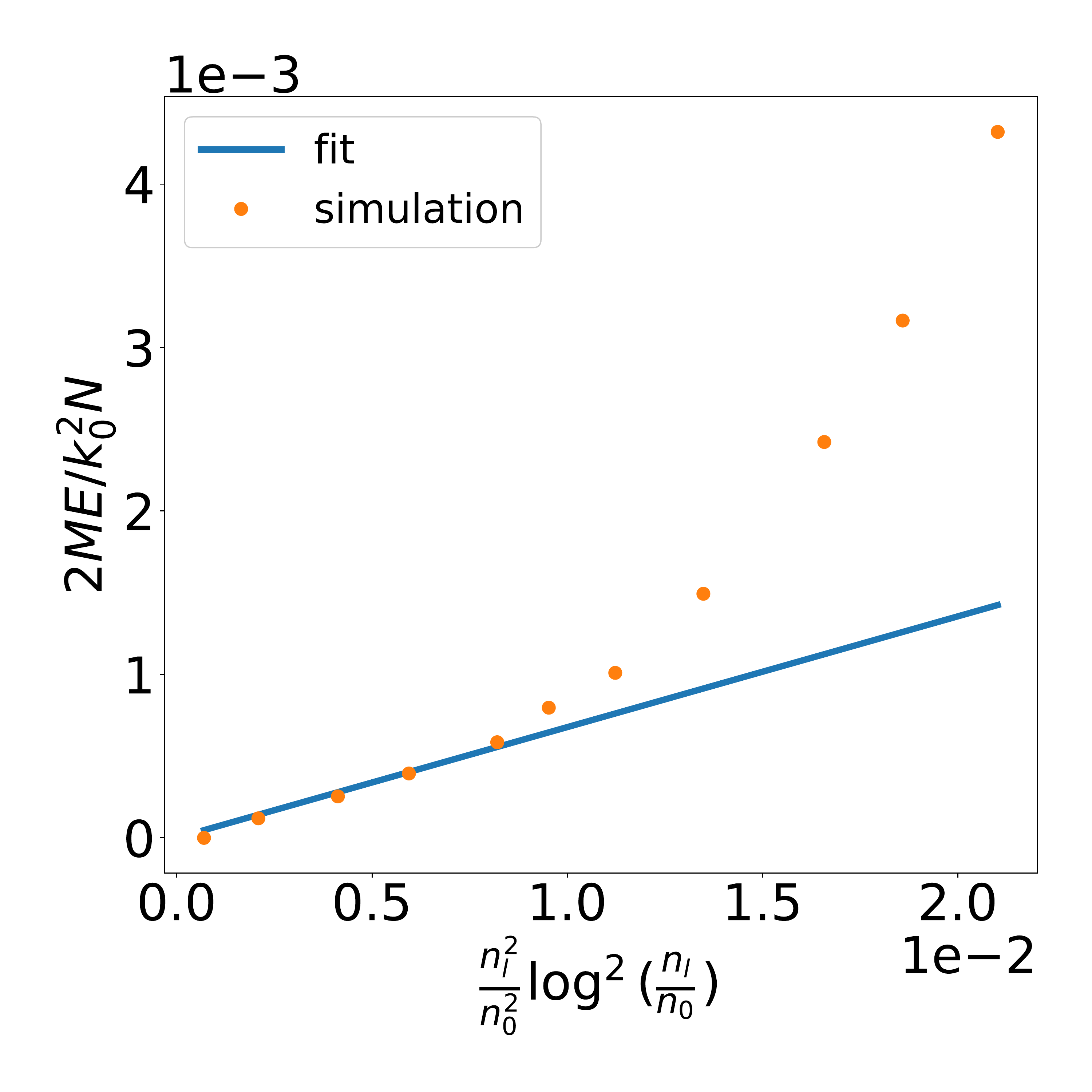}
    \begin{picture}(0,0)
    {\def\unitlength{}
    \put(0.16\textwidth,0.4\textwidth){\includegraphics[height=0.4\textwidth]{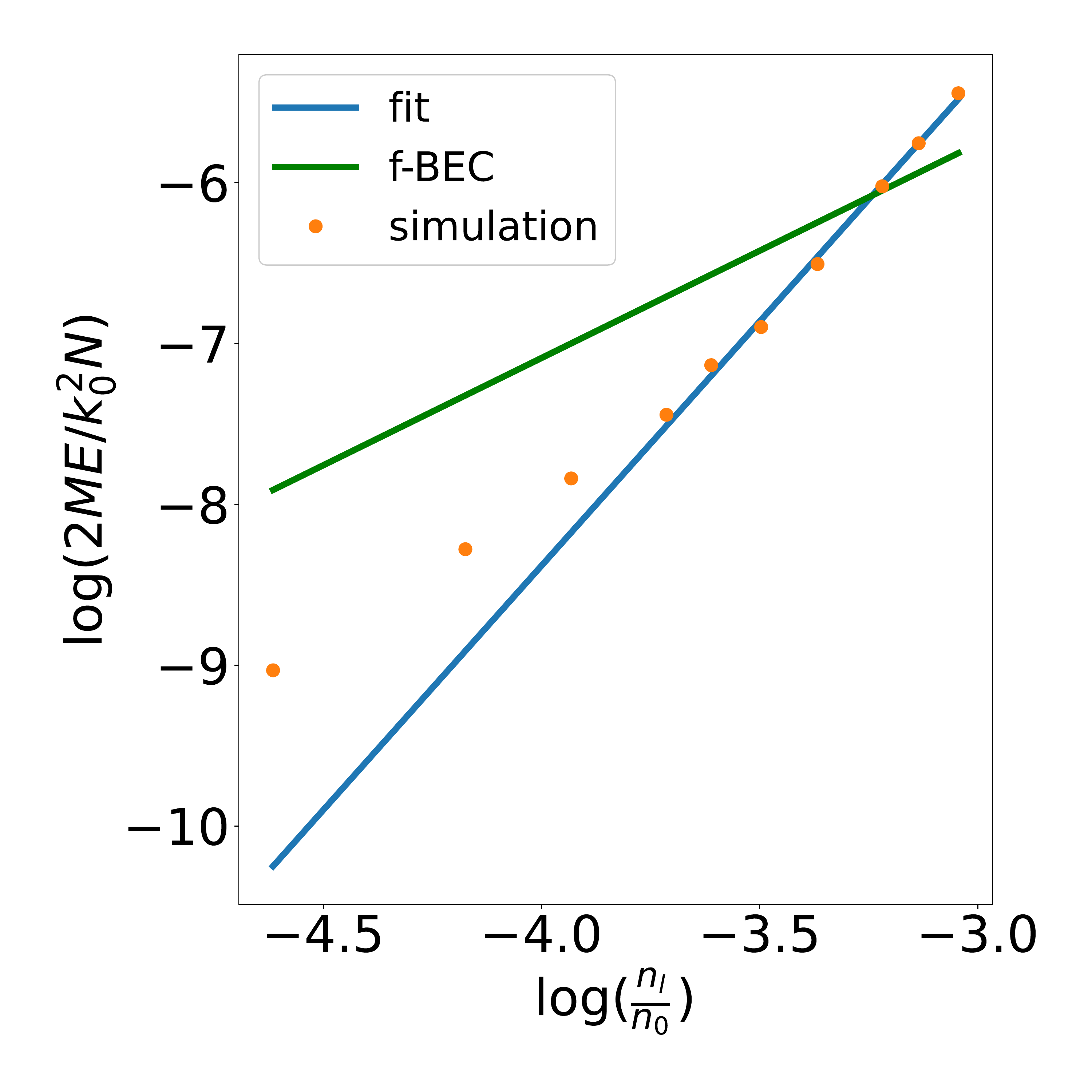}}
    }
    \end{picture}
    \end{minipage}
    \caption{The scaling behavior of the energy \cref{Esum} as a function of discrete densities $n_l$ given by \cref{nl}. The relative uncertainty is of the order $10^{-3}$, which is not indicated in the figures.
    The main figure shows the linear dependence of per-particle energy on $n^2 ln^2(n)$ at low density.
    The inset shows the high density scaling $E \sim n^{3.0}$.
    The green line is the energy of the f-BEC state. The zero density energy constant is set to zero.
    }
    \label{fig:MC}
\end{figure}

\Cref{El} allows us to perform MC simulation with the metropolis algorithm, where $z, z_2, ..., z_N$ are sampled according to the distribution $\Psi_F^{(l)*}(z) \Psi_F^{(l)}(z)$ and $w$ is sampled according to $p(w) = \frac{1}{2\pi |w-z|} e^{-|w-z|} dw = \frac{1}{2\pi} e^{-r} dr d\theta$.
To accomplish the simulation, $10^{8}$ MC loops are run to thermalize, then $10^{10}$ MC loops are run, and measurements are performed every $10^{2}$ MC loops.
The whole process is discussed in \ref{apx_MC}.

The MC simulation is performed with
the number of particles $N = 60$, $l$ is chosen to be
$l = 10, 11, 12, 14, 16, 18, 20, 25, 32, 50, 100$, which determines the density through \cref{nl}.
In the inset of \cref{fig:ratiok}, the averaged momentum $\langle |k| \rangle$ and averaged squared momentum $\langle k^2 \rangle$ are fitted with second-order polynomials. To see the low density analytical relation \cref{linearTerm}, one needs to fit $\langle |k| \rangle$ and $\langle |k|^2 \rangle$ with only low density points, \textit{i.e.}, fitting with only points in range $[0, n^{\rm (max)}]$. Then the ratio of their fitted linear term coefficients at low densities approaches a constant asymptote of $2$, as shown in the main figure of \cref{fig:ratiok} (the figure shows a logarithmic plot, for which the predicted ratio corresponds to the situation when the logarithm goes to $1$).

The averaged energy shown in \cref{fig:MC} scales as $\langle E \rangle \sim n^2 \ln^2n$ at $l > 20$ with the corresponding density $n < n^*_1 \equiv \frac{k^2_0}{82 \pi}$.
In the experimental setup of \cite{bracamontes2022realization}, $k_0 \approx \frac{1}{5} \frac{2\pi}{\lambda}$ has a corresponding density scale of $n \lessapprox \frac{6.13 \cross 10^{-3}}{\lambda^2}$, or the filling fraction of about $\sim \frac{1}{40}$.
When comparing these results with the f-BEC energy $\frac{2ME}{k_0^2 N} = \frac{2}{(2\pi)^{4/3}} (\frac{n}{n_0})^{4/3}$\cite{gopalakrishnan2011universal,hsieh2022helical} ($n_0 = k_0^2 / 2\pi$, which is consistent with \cref{nl}), the variational energy of the CSL state is lower than the f-BEC energy at $l > 12$, which indicates the CSL state is stable up to higher density $n^*_2 \equiv \frac{k^2_0}{50 \pi}$ for the dimensionless interaction $g\sim 1$. This allows observation of the CSL state at a relative higher density in experiments, $n \lessapprox \frac{1.0 \cross 10^{-2}}{\lambda^2}$, or about $\frac{1}{25}$ filling fraction observable in time of flight experiments.
In this region, $E \sim n^3$. We attribute the deviation from $n^2 \ln^2(n)$ to the non-uniformity CSL\cite{sedrakyan2015spontaneous,wang2022emergent}, where the CDW order or, alternatively, out of plane Ising order, is shown to coexist with the CSL state lowering the overall energy of the CSL state. A similar scaling behavior can also emerge in a 1D Fermi gases with cubic dispersion, \textit{i.e.}, the edge excitation of Laughlin states\cite{dispersion1994giovanazzi,li2021dynamics}.
A low energy effective hydrodynamic description of the edge excitations indicates the $E \sim n^3$ can be captured by an emergent surface tension term.
Note that at high particle density, tensor network calculation of the lattice CS representation also predicts nonuniform CSL\cite{wang2022emergent,sedrakyan2017topological,wang2018chern,wang2022chern}.

It is also interesting to compare the CSL energy with the BEC energy $\frac{2ME}{k_0^2 N} = g \pi n / n_0$, as a result of which, the weakly interaction bosons, $g \ll (n/n_0) \ln^2(n/n_0)$, with moat-like dispersion remain Bose-Einstein condensed. The phase diagram \cref{fig:PD} is obtained based on this comparative analysis.

\section{Conclusion and outlook}
We performed variational a Monte Carlo analysis of the equation of state corresponding to the CSL state of strongly interacting bosons with the moat dispersion. We found the density interval where the chemical potential scales with the density as $\mu_{CSL}\sim n^2\ln^2(n)$. Numerical data shows a high precision fit up to density $\frac{k^2_0}{82 \pi}$ beyond which the scaling gradually changes. In the intermediate density region $\frac{k^2_0}{82 \pi} < n < \frac{k^2_0}{50 \pi}$, which indicates the non-uniform CSL with surface tension low energy effective description. Combined with previous analytical and numerical results, the phase diagram \cref{fig:PD} is proposed, where the BEC phase dominates the weakly interacting region. In contrast, in the strongly interacting region, the emergent CS field leads to the CSL phase. The finite size scaling of the energy in the emergent CSL phase remains an exciting and open problem.
To experimentally identify such a CSL state unambiguously, one needs a method to access the information of the collective modes in the whole system. Although some experimental techniques, such as polarized neutron scattering\cite{simonet2012magnetic} or Raman susceptibility measurement\cite{wang2020the}, may provide indirect evidence of the time-reversal symmetry breaking and nonlocal chirality.

Flatband lattice systems, similar to moat band, also give rise to the CSL state with emergent lattice CS field\cite{maiti2019fermionization}. It is also shown that a fermionic flatband system with the disorder can harbor compressible non-Fermi liquid or strange metal with chaos\cite{wei2021optical,wei2021disordered}. How the moat band system behaves in the presence of disorder still remains an open problem. The moat band becomes flat in the $m\to \infty$ limit, enhancing the interaction effect. So the $\Psi_F$ in the wavefunction \cref{PhiB} could acquire a more complicated structure than a simple Slater determinant. Localization effects are also predicted in the disordered CSL\cite{kao2021disorder}. Notably, the limit $k_0 \to 0$ leads to the Bose glass as observed in \cite{fisher1989boson,damski2003atomic}.

\section*{Acknowledgments} The discussions with A. Kamenev, J. Radic, J. Zhang, and  V. Galitski, at the early stage of this work, are acknowledged with thanks.  We also acknowlegdge the communication from E. Lake and thank him for inviting our attention to a relevant reference. 

\appendix
\section{Proof of equation \ref{absk2int}} \label{apx_proof_absk2int}

We start by observing the integral representation of $|k|$
\begin{equation} \label{abs_int}
    |k| = \frac{1}{\sqrt{\pi}} k^{\dagger}k \int_{-\infty}^{\infty} ds e^{-s^2 k^{\dagger}k}.
\end{equation}

Using the fact that $e^{-s^2 \bar{\partial} \partial}$ acts as a translation generator
\begin{equation}
\begin{aligned}
    &\int_{-\infty}^{\infty} ds e^{-s^2 k^{\dagger}k} f(z, \bar{z}) \\
    =& \int_{-\infty}^{\infty} ds e^{4s^2 \bar{\partial} \partial} f(z, \bar{z}) \\
    =& \int_{-\infty}^{\infty} ds \int dw d\bar{w} \frac{1}{\pi} e^{-2s w \partial} e^{-2s \bar{w} \bar{\partial}} e^{-\bar{w} w} f(z, \bar{z}) \\
    =& \int_{-\infty}^{\infty} ds \int dw d\bar{w} \frac{1}{\pi} e^{-\bar{w} w} f(z-2sw, \bar{z}-2s\bar{w}) \\
    =& \int_{-\infty}^{\infty} ds \int dw d\bar{w} \frac{1}{4\pi s^2} e^{-(\bar{z} - \bar{w}) (z - w) / 4s^2} f(w, \bar{w}) \\
    =& \int dw d\bar{w} \frac{1}{2 \sqrt{\pi}} \frac{1}{|z-w|} f(w, \bar{w}).
\end{aligned}
\end{equation}

Combining the equations above, one arrives at \cref{absk2int} in the main text.

\section{Derivation of scaling behavior \ref{scaling:k}} \label{apx_proof_scaling:k}
As what is done in the main text, the bosonic average can be replaced by the fermionic average with an extra gauge field $A(z) = -i \sum_j (z-z_j) / |z-z_j|^2$. After using the integral representation of the modulus \cref{abs_int} and decompose $|k - A|^2$ into mean-field part ad the fluctuation, one gets
\begin{equation} \label{apx_k}
\begin{aligned}
    \langle |k| \rangle_B &= \langle |k - \bar{A}| \rangle_F \\
    &= \frac{1}{\sqrt{\pi}} \langle |k - A|^2 \int_{-\infty}^{\infty} ds e^{-s^2 |k - A|^2} \rangle_F\\
    &= \frac{1}{\sqrt{\pi}} \langle (|k - \bar{A}|^2 + H_0) \int_{-\infty}^{\infty} ds e^{-s^2 (|k - \bar{A}|^2 + H_0)}\rangle_F.
\end{aligned}
\end{equation}

The Baker-Campbell-Hausdorff formula can be employed to further simplify the exponential part
\begin{equation}\label{BCH}
    e^{-s^2 (|k - \bar{A}|^2 + H_0)} \approx e^{-s^2 H_0 + 2 s^4 [ \bar{\partial} \partial, H_0 ]} e^{-s^2 |k - \bar{A}|^2}.
\end{equation}

As before, under the average, the mean-field part is constant, \textit{i.e.}, $k - \bar{A} \sim k_0$. And $e^{-s^2 H_0 + 2 s^4 [ \bar{\partial} \partial, H_0 ]} e^{-s^2 k_0^2}$ can be Taylor expanded then integrated over $s$
\begin{equation}\label{Taylor}
    \frac{1}{\sqrt{\pi}} \int_{-\infty}^{\infty} ds e^{-s^2 (|k - i\bar{A}|^2 + H_0)} \sim \frac{1}{k_0} - \frac{1}{2k_0^3}H_0 + \frac{3}{2k_0^5} [ \bar{\partial} \partial, H_0 ] + \frac{3}{4k_0^5}H_0^2.
\end{equation}

Terms ignored in \cref{BCH,Taylor} are of higher order in density $nln(n)$, therefore these equations hold for low density.

By dimensional analysis, one can evaluate the scaling of terms in \cref{Taylor}
\begin{equation} \label{n2scaling}
\begin{aligned}
    \langle [ \partial^2, H_0 ] \rangle_F \sim D_1 n^2 ln^2(n), \\
    \langle H_0^2 \rangle_F \sim D_2 n^2 ln^2(n),
\end{aligned}
\end{equation}
where $D_1$ and $D_2$ are constant.

Inserting the scaling \cref{H0_scaling,n2scaling} into \cref{apx_k,Taylor}, one gets \cref{scaling:k}
\begin{equation}
    \langle |k| \rangle_B = k_0 + \frac{C}{2k_0} n ln(n) + \frac{D'}{k_0^3} n^2 ln^2(n) + O(n^3 ln^3(n)),
\end{equation}
where $C$ is the same constant coming from \cref{H0_scaling} and $D'$ is a constant depending on $C$, $C$, $D_1$ and $D_2$.

\section{Brief discussion of MC simulation for \cref{El}} \label{apx_MC}

\begin{figure}[!ht]
\centering
    \begin{tikzpicture}[node distance=2cm]
    
    \node (start) [startstop] {Start};
    
    \node (in) [io, below of=start, yshift=0.7cm] {Randomly initialize $z$, $\left. z_i \right|_{i=2,3,\cdots,N}$ (and $w$ if evaluating $\langle |k| \rangle$)};

    \node[text width=10cm] (pro1) [process, below of=in, yshift=0.4cm] {Pick one of the $z$ or $\left. z_i \right|_{i=2,3,\cdots,N}$ and propose new $z$ or $z_i$ according to distribution $p(z_{\rm old}, z_{\rm old}) = \frac{1}{2\pi} e^{-|z_{\rm old}-z_{\rm new}|^2/2}$\\
    (or $w$ from $p(w) = \frac{1}{2\pi |w-z|} e^{-|w-z|}$)};
    
    \node (dec1) [decision, below of=pro1, yshift=0.3cm] {condition 1};
    
    \node (pro2) [process, below of=dec1, yshift=0.5cm] {Update the proposed $z$ or $\left. z_i \right|_{i=2,3,\cdots,N}$ (or $w$)};

    \node (dec2) [decision, below of=pro2, yshift=0.6cm] {condition 2};

    \node (pro3) [process, below of=dec2, yshift=0.5cm] {Compute the $\mathcal{A}$/$\mathcal{B}$};

    \node (dec3) [decision, below of=pro3, yshift=0.6cm] {condition 3};
    
    \node (out) [io, below of=dec3, yshift=0.5cm] {Compute average of the $\mathcal{A}$/$\mathcal{B}$ and hence the $\langle |k| \rangle$, $\langle |k|^2 \rangle$ and $E$};
    
    \node (stop) [startstop, below of=out, yshift=0.7cm] {End};
    
    \draw [arrow] (start) -- (in);
    \draw [arrow] (in) -- (pro1);
    \draw [arrow] (pro1) -- (dec1);
    \draw [arrow] (dec1) -- node[anchor=east] {yes} (pro2);
    \coordinate[right of = dec1] (aux1);
    \draw [arrow] (dec1.east) -| node[anchor=south] {no} (aux1)++(2cm,0) |- (dec2.north);
    \draw [thick] (aux1)++(2cm,0) -- (aux1);
    \draw [arrow] (pro2) -- (dec2);
    \draw [arrow] (dec2) -- node[anchor=east] {yes} (pro3);
    \coordinate[right of=dec2] (aux2);
    \draw [thick] (dec2.west)++(-4.4cm,0) -- node[anchor=south] {no} (dec2.west);
    \draw [arrow] (pro3) -- (dec3);
    \draw [arrow] (dec3) -- node[anchor=east] {yes} (out);
    \coordinate[left of=dec3] (aux3);
    \draw [arrow] (dec3.west) -| node[anchor=south] {no} (aux3)++(-4cm,0) |- (pro1.west);
    \draw [thick] (aux3)++(-4cm,0) -- (aux3);
    \draw [arrow] (out) -- (stop);
    
    \end{tikzpicture}
    \caption{Flowchart of the MC algorithm. \\
    condition 1: $r < \left| \frac{\Psi_F^{(l)*}(z_{\rm new}) \Psi_F^{(l)}(z_{\rm new})}{\Psi_F^{(l)*}(z_{\rm old}) \Psi_F^{(l)}(z_{\rm old})} \right|^2$ where $r$ is a random number uniformly taken from $[0,1]$ (newly proposed $w$ is always accepted)\\
    condition 2: In thermalization \\
    condition 3: All loops end
    }
    \label{fig:flowchart}
\end{figure}

The two integrals in \cref{El} can be evaluated separately
\begin{equation} \label{EA}
    E_A = \frac{k_0^2}{2M} \int dz d\bar{z} \prod_{i=2}^{N} dz_i d\bar{z}_i \Psi_F^{(l)*}(z) \Psi_F^{(l)}(z) \mathcal{A},
\end{equation}
and
\begin{equation} \label{EB}
    E_B = \frac{k_0^2}{2M\pi} \int dw d\bar{w} \int dz d\bar{z} \prod_{i=2}^{N} dz_i d\bar{z}_i {\frac{e^{-|w-z|}}{2\pi |w-z|}} \Psi_F^{(l)*} (z) \Psi_F^{(l)} (z) \mathcal{B},
\end{equation}
where
$\frac{E_l}{N} = \frac{k_0^2}{2M} + E_A - E_B$

One can view \cref{EA,EB} expectation value of $\mathcal{A}$ or $\mathcal{B}$ with respect to probability distributions
\begin{equation}
    p(\{z_i\},w) = \frac{e^{-|w-z|}}{2\pi |w-z|}
    \Psi_F^{(l)*} (z) \Psi_F^{(l)} (z) 
\end{equation}
And then, the standard Metropolis procedure can be used to evaluate the integrals, the algorithm of which is illustrated in \cref{fig:flowchart}.

\bibliographystyle{elsarticle-num} 

\end{document}